\documentclass[useAMS,usenatbib]{mn2e}

\usepackage{amsmath,fleqn,graphicx,txfonts}
\usepackage{epsf}
\usepackage{color}
\usepackage[dvips]{epsfig}

\begin{document}

\label{firstpage}

\title[Formation scenario for Scl-dE1 GC1]{Star Cluster collisions -
  a formation scenario for the Extended Globular Cluster Scl-dE1 GC1}

\author[Assmann et al.]{
  P. Assmann$^{1}$ \thanks{E-mail: passmann@astro-udec.cl},
  M.I. Wilkinson$^{2}$ \thanks{miw6@astro.le.ac.uk},
  M. Fellhauer$^{1}$ \thanks{mfellhauer@astro-udec.cl},
  R. Smith$^{1}$ \thanks{rsmith@astro-udec.cl} \\
  $^{1}$ Departamento de Astronom\'{i}a, Universidad de
  Concepci\'{o}n,
  Casila 160-C, Concepci\'{o}n, Chile \\
  $^{2}$ Department of Physics \& Astronomy, University of Leicester,
  University Road, Leicester LE1 7RH, UK}

\pagerange{\pageref{firstpage}--\pageref{lastpage}} \pubyear{2010}

\maketitle

\begin{abstract}
  Recent observations of the dwarf elliptical galaxy Scl-dE1 (Sc22) in
  the Sculptor group of galaxies revealed an extended globular cluster
  (Scl-dE1~GC1), which exhibits an extremely large core radius of
  about $21.2$~pc.  The authors of the discovery paper speculated on
  whether this object could reside in its own dark matter halo and/or
  if it might have formed through the merging of two or more star
  clusters.  In this paper, we present N-body simulations to explore
  thoroughly this particular formation scenario.  We follow the merger
  of two star clusters within dark matter haloes of a range of masses
  (as well as in the absence of a dark matter halo).  In order to
  obtain a remnant which resembles the observed extended star cluster,
  we find that the star formation efficiency has to be quite high
  (around 33~per cent) and the dark matter halo, if present at all,
  has to be of very low mass, i.e.\ raising the mass to light ratio of
  the object within the body of the stellar distribution by at most a
  factor of a few. We also find that expansion of a single star
  cluster following mass loss provides another viable formation
  path. Finally, we show that future measurements of the velocity
  dispersion of this system may be able to distinguish between the
  various scenarios we have explored.
\end{abstract}


\section{Introduction}
\label{sec:intro}

The general understanding of low luminosity stellar systems is that
they come in two distinct flavours. On the compact side we have the
globular clusters whose properties are consistent with a purely
stellar composition with no need to invoke dark matter (DM) to explain
their dynamical masses~\citep[e.g.][]{Lane2010}.  On the more extended
side, we have the dwarf spheroidal galaxies (dSph) which have similar
luminosities to globular clusters but are much more extended and show
enhanced velocity dispersions.  These entities are believed to be
highly DM dominated, typically having estimated mass-to-light (M/L)
ratios of several 10s to 100s~\citep[e.g.][]{mat98,wal09}, with
values as high as 1000 being claimed for some of the faintest
systems~\citep[e.g.][]{Strigari2008,Kleyna2005}. These high M/L values
make the dSphs the most DM dominated objects known in the local
universe~\citep[see e.g.][for a convenient review]{mat98}.  Between
these two regions of parameter space there seems to be a clear
gap~\citep{gil07} in the distribution of effective radii.  This gap is
visible over a wide range of absolute magnitudes, i.e.\ for several
orders of magnitude in stellar mass.

With the discovery of the new ultra-faint galaxies around the Milky
Way~\citep[see e.g.][and references therein, for the latest
discoveries]{bel10}, there seems to be a closing of the gap at very
low stellar masses, which could be due to the tidal disruption of
those small systems as they orbit the Milky Way. In this region, the
distinction between star clusters and dwarf galaxies becomes blurred -
in the absence of detailed chemical information about the stellar
populations and star formation history, we can not be sure whether a
particular object is a dissolving star cluster or a disrupting
dwarf galaxy.

But this faint, low-mass end is not the only region of the diagram
where the two distinct populations of stellar systems come closer to
each other, reducing the gap between their half-light radius
distributions.  At the high-luminosity end, we have the population of
ultra-compact dwarf (UCD) galaxies, which have masses up to
$10^{8}$~M$_{\odot}$ and effective radii of $20$--$40$~pc.  Their
formation mechanism is still under debate.  Do we see some freakish,
high-mass outliers of the globular cluster
mass-function~\citep[e.g.][]{mie02}?   Are they the product of many
star clusters, which have formed in a dense star forming region and
then subsequently merged~\citep{fel02}?  Or are they instead the
stripped nuclei of dwarf galaxies~\citep{bek01}?  They clearly fall on
the star cluster side of the above-mentioned gap but are more extended
than typical clusters, suggesting a relation to dwarf galaxies.
Further, some of them show evidence of M/L-ratios which can not be
explained by simple stellar population models~\citep{has05}.  The jury
is still out on whether or not UCDs have a dynamically significant DM
component.

Even in the `normal' range of globular cluster luminosities, we now
know of objects which are more extended than the regular globular
cluster family.  Regular globular clusters have effective radii of
order a few pc.  However, there is also a population of clusters with
effective radii in the range $10$--$20$~pc.  We see those extended
globular clusters (EGC) as 'faint fuzzy' star clusters in the discs of
S0 galaxies~\citep{lar00} and as Huxor objects in the Halo of
M31~\citep{hux05}.

Extended globular clusters were first found by \citet{hux05} in
M31. The authors searched for old objects in the halo of Andromeda and
found objects which are similar to globular clusters in colour and
luminosity, but with half-light radii of more than $30$~pc. Their
relative distances to the centre of M31 are between $15$ and
$35$~kpc. Similarly, \citet{hwa05} found the first extended globular
clusters in the halo of the Local Group dwarf irregular galaxy NGC
$6822$. The authors reported that the projected distances of these
EGCs from the centre of NGC $6822$ are close to $13$~kpc, and
estimated their half-light radii to be larger than $20$~pc. This kind
of object was previously unknown and could imply the existence of an
intermediate population of objects between usual GCs and dSph
galaxies. Further discoveries have been added to the original
observations~\citep{Huxor2008}, but, it is still unclear whether these
are really a distinct population, or rather the large effective radius
tail of a continuous distribution of cluster sizes.

Meanwhile, EGCs have also been found in many other galaxies.  Extended
globular star clusters (EGC) are clusters of stars for which the
half-light radii ($r_{\rm h}$) exceed $8-10$~pc.  Their occurrence
seems to be aleatory: they represent less than $3\%$ of the clusters
in NGC 5128~\citep{gom07}, $9\%$ of the Milky Way
clusters~\citep{har96}, and are more than $20\%$ of the clusters in
M51~\citep{cha04}.  Even though EGCs are usually found at large
galactocentric radii, they can also be observed in dwarf
galaxies~\citep{cos09}.  This additional dichotomy within the star
cluster population is also not well understood.  As in the case of
UCDs, a possible scenario for their formation is again the merging of
two or more star clusters, which have formed in a confined
region~\citep{bru09}.

In this work we focus on the first EGC found in a dwarf elliptical
galaxy. In the Sculptor group of galaxies, \citet{cos09} found an EGC
(GC1) around the dwarf elliptical galaxy Scl-dE1 (Sc22).  It has a
half-light radius of $21.8$~pc and an absolute magnitude of M$_{\rm
  V}=-6.7$. Another reported characteristic of this EGC is that its
stellar population seems to be indistinguishable from the population
of the parent dwarf.  By considering a king profile \citep{kin62} for
the V-band surface brightness profile of this EGC, the authors were
able to measure a core radius of $21.2$~pc, a concentration index
c$=0.65$ and a central surface brightness of $23.1$~V mag
arcsec$^{-2}$. {\color{red}Observationally, the half-light radius is
  defined as the projected radius at which the surface brightness has
  dropped to half its central value. In the theoretical literature,
  the term core radius is used to refer to the natural scalelength of
  the model under consideration, for example, the King model. It is
  important to note that these definitions do not, in general coincide
  and in real systems their temporal evolution may be qualitatively
  different due to processes such as mass segregation of stellar
  remnants with high mass-to-light ratios~\citep[see
    e.g.][]{Hurley2007,Wilk2003}. In this paper, we estimate the core
  radii of our models by fitting King profiles to them in the same
  manner as in \citet{cos09}. Thus, our quoted core radii can be
  compared directly to that of Scl-dE1 GC1 reported in that
  paper. }
 
By means of numerical simulations we want to investigate whether (1)
this EGC could have formed in a DM halo of its own (a scenario in
which all globular clusters may reside in their own DM halo was
proposed by \citet{mas05}); (2) this EGC could be the product of the
merging of two (or more) star clusters. We combine these two scenarios
and simulate the merger of two star clusters orbiting in a DM halo.  A
similar scenario has been proposed by \citet{ass10} as a possible
formation mechanism for the luminous components of dwarf galaxies such
as dSphs.

In Section~\ref{sec:setup} we describe the setup of our simulations in
detail.  In Section \ref{sec:res} we present and discuss our main
results, namely that no additional DM is needed to explain the
properties of GC1. Finally, in Section \ref{sec:conc} we summarise our
conclusions.

\section{Setup}
\label{sec:setup}

Initially, we place two star clusters at a distance of $0.5$~kpc from
each other, at $x=\pm 0.25$~kpc.  While this choice is arbitrary, it
ensures that the two clusters are well-separated at the beginning of
our simulations.  Each star cluster is represented by a Plummer sphere
with a Plummer radius (effective radius) of $11$~pc and is initially
truncated beyond a cut-off radius of $44$~pc, {\color{red}because less
  than two per cent of the cluster mass resides beyond this radius. }
The clusters begin the simulations in equilibrium, but we then mimic
the effect of early gas expulsion by artificially reducing the mass of
the individual star particles during the first crossing time of the
star cluster ($10$~Myr). {\color{red}This mass-loss represents not
  only the loss of gas which has not formed stars, but also mass-loss
  from the rapid evolution of massive stars, although we note that we
  do not use a sophisticated mass-loss algorithm as in \citet{dab10},
  and in the subsequent evolution we do not take mass-loss due to
  stellar evolution into account. The possible impact of mass-loss
  from stars on the late-time evolution of a merger remnant at the
  centre of a dark matter potential well is itself an interesting
  issue to be explored elsewhere.}

The initial mass of the star clusters is varied according
to the adopted star formation efficiency (SFE) in order to obtain
final masses (after gas expulsion) of $2.5 \times 10^{4}$~M$\odot$ in
all simulations.  We consider three values for the SFE, namely $10$,
$33$ and $100$~per cent which lead to initial masses of our star
clusters of $2.5 \times 10^{5}$, $7.6 \times 10^{4}$ and $2.5 \times
10^{4}$~M$_{\odot}$, respectively. Each cluster is represented by
$10^5$~particles.

Our model clusters are placed inside spherical DM haloes of different
masses and density profiles. {\color{red}As we are primarily interested
  in the early evolution of the merger remnant, we treat the halo as
  an isolated DM halo, ie.we do not simulate the dark matter halo of
  the dwarf elliptical galaxy Scl-dE1 (Sc22).} We use either a
cored Plummer profile~\citep{plu11,aar74} with a scale length of
$500$~pc or a cusped NFW profile~\citep{nav97,deh05} with a
characteristic radius of $500$~pc.  The mass of the halo enclosed
within $500$~pc ($M_{\rm DM}$) is varied over many orders of magnitude
namely: none, $5.6 \times 10^{4}$, $5.6 \times 10^{5}$, $5.6 \times
10^{6}$, $5.6 \times 10^{7}$ and $5.6 \times 10^{8}$~M$_{\odot}$.  The
halo is modeled using 1,000,000 particles.  As a cut-off radius for
our halo we take $2.5$~kpc in the Plummer cases and the virial radius
for the NFW profiles.  Due to the restrictions of fixed scale-length
and fixed enclosed mass, the concentrations of our NFW models vary
from $2.1$ up to $89.5$. In Table~\ref{tab:halo}, we give an overview
of the main parameters of the dark matter haloes we choose.

\begin{table}
  \centering
  \caption{Initial parameters of the dark matter haloes used.  The
    first column denotes the halo profile, which is either Plummer (P)
    or NFW (N).  The second column gives the mass enclosed within
    $500$~pc.  In the third column we give the scale length of the
    halo, either the Plummer radius or the scale radius of the NFW
    profile.  The fourth column gives the cut-off radius of the halo
    which is chosen to be either $5$~R$_{\rm pl}$ or the virial radius
    of the NFW halo.  Finally, the last column denotes the circular
    speed of the halo at the initial position of the star clusters.}
  \label{tab:halo}
  \begin{tabular}{crrrr}
    P/N & $M_{\rm DM}$ & $r_{\rm s}$ & $r_{\rm c}$ &
    $v_{\rm c}(250\,{\rm pc})$ \\
    & [M$_{\odot}$] & [kpc] & [kpc] & [km\,s$^{-1}$] \\ \hline
    P & $5.6 \times 10^{4}$ & 0.50 & 2.50 & 0.17 \\
    N & $5.6 \times 10^{4}$ & 0.50 & 1.05 & 0.60 \\
    P & $5.6 \times 10^{5}$ & 0.50 & 2.50 & 0.55 \\
    N & $5.6 \times 10^{5}$ & 0.50 & 3.03 & 1.90 \\
    P & $5.6 \times 10^{6}$ & 0.50 & 2.50 & 1.75 \\
    N & $5.6 \times 10^{6}$ & 0.50 & 7.81 & 6.00 \\
    P & $5.6 \times 10^{7}$ & 0.50 & 2.50 & 5.52 \\
    N & $5.6 \times 10^{7}$ & 0.50 & 18.98 & 18.97 \\
    P & $5.6 \times 10^{8}$ & 0.50 & 2.50 & 17.45 \\
    N & $5.6 \times 10^{8}$ & 0.50 & 44.73 & 59.98 \\ \hline
  \end{tabular}
\end{table}

In the case of the very low mass halo ($5.6 \times 10^{4}$~M$_{\odot}$)
we also consider different initial relative velocities for the star clusters.
The velocities are shown in Table.~\ref{tab:vel}.  In the first case
the clusters are initially at rest.  In the second and third case they
have transverse velocities, while in the last case, they have
velocities directed towards each other.  The choice of velocities is
arbitrary, but we show later that our results, to first order, do not
depend strongly on this choice of the initial velocities.

\begin{table}
  \centering
  \caption{Initial relative velocities of the star clusters in km\,s$^{-1}$.}
  \label{tab:vel}
  \begin{tabular}{cccc} \hline
    case & $v_{x}$ & $v_{y}$ & $v_{z}$ \\ \hline
    1 & $0.00$ & $0.00$ & $0.00$ \\
    2 & $0.00$ & $\pm 0.10$ & $0.00$ \\
    3 & $0.00$ & $\pm 0.25$ & $0.00$ \\
    4 & $\pm 0.25$ & $0.00$ & $0.00$ \\ \hline
  \end{tabular}
\end{table}

We also perform a different set of simulations where we use an initial
model for the two star clusters which has a scale-radius
(Plummer-radius) of only $4$~pc.  This is a more standard value for
young star clusters \citep[see e.g.][and many more]{whi99}.

Finally, we show a suite of simulations in which we place a single
star cluster in the centre of the halo and follow its evolution,
without a merger, due to the expansion produced by a low SFE.

To perform our simulations we use the particle-mesh code {\sc
  Superbox} ~\citep{fel00}. This code has many advantages.  One of
them is that we can use an arbitrarily high number of particles to
model the cluster and the dark matter. {\color{red}Two-body effects are suppressed
in a code of this type and only the smoothed potential is taken into
account. Assuming that the simulation results in a merged object which
resides in the centre of the DM halo, thereby having a large
relaxation-time (of order a Hubble time or longer), this is an appropriate
approximation since we would expect that binary formation and two-body encounter
would have only a minor effect on our results.}

Another advantage is that {\sc Superbox} allows us to
set different levels of resolution for the different levels of
high-resolution grids used in the code.  Therefore the code provides
the correct resolution in the places where it is most needed.  For our
models, we use resolutions in the highest resolution grid of $16$~pc
for the DM halo and $0.4$~pc for the star clusters.  This means that
the internal forces within a star cluster, the forces between the two
star clusters when they interact and the forces from the star clusters
acting on the halo have very high resolution, and the overall force of
the background potential of the halo has sufficient resolution to
resolve the orbits of the star clusters and the tidal forces acting on
them, while keeping computational costs low.

In this code, two-body effects, like binary
formation and two-body relaxation are neglected, since they do not play
an important role if the final object resides in a DM potential.

\section{Results}
\label{sec:res}

\begin{figure}
  \begin{center}
    \epsfxsize=6.1cm
    \epsfysize=6.1cm
    \epsffile{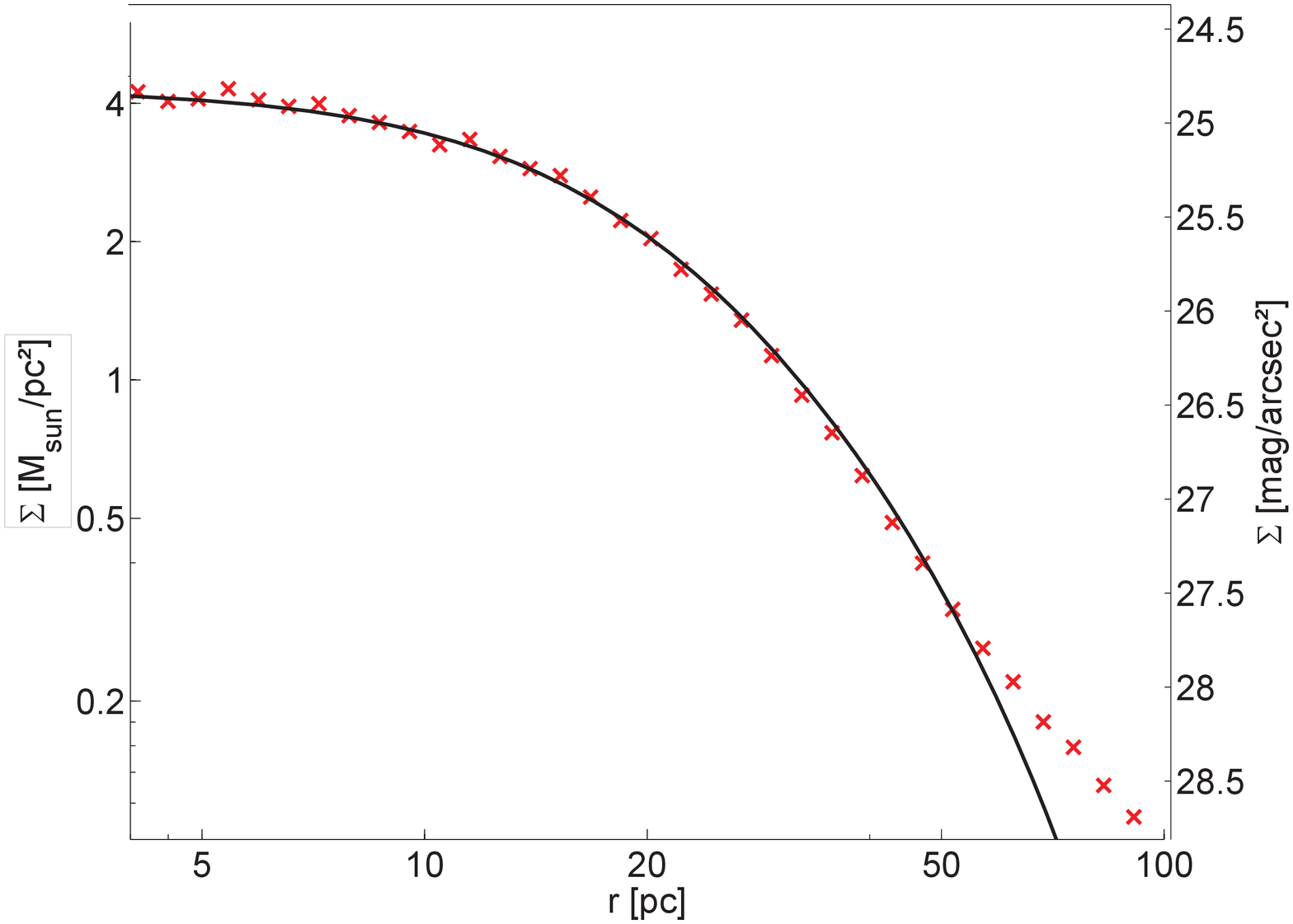}
    \epsfxsize=6.1cm
    \epsfysize=6.1cm
    \epsffile{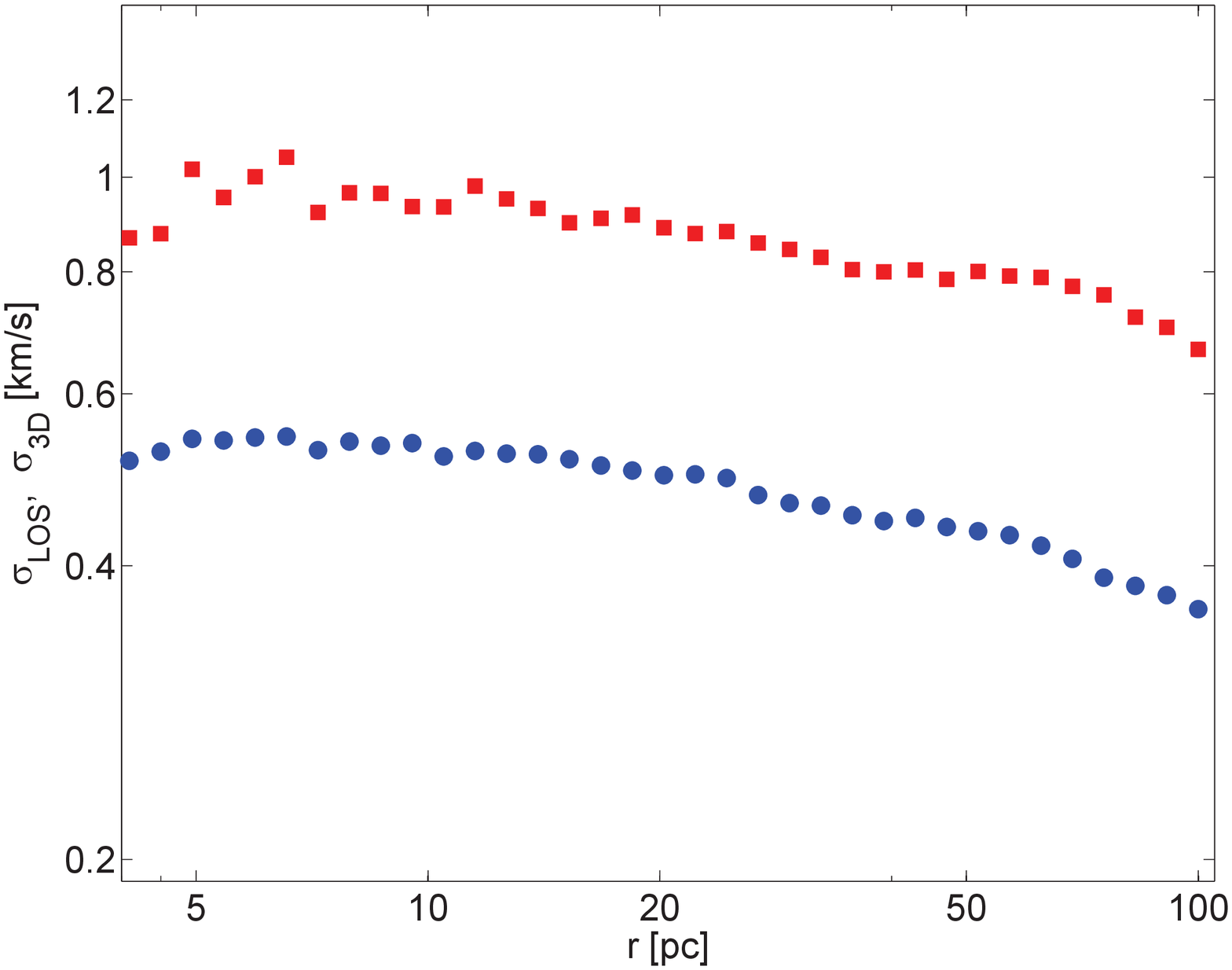}
    \caption{Plots showing the properties of the remnant resulting
      from a cluster merger within a low mass cusped (NFW) DM halo of
      mass $M_{\rm DM}=5.6 \times 10^{4}$~$[$M$_{\odot}]$ and SFE for
      the star clusters of $33\%$ (see Table~\ref{tab:res2}).  The
      surface density profile of the EGC is shown in the top panel.
      The (red online) crosses are the data points, and the solid
      curve is the King profile fit.  It has a core radius $r_{\rm
        c}=21.5$~pc and a tidal radius $r_{\rm t} = 196$~pc.  In the
      bottom panel the velocity dispersions are shown.  The (blue
      online) circles are the line-of-sight velocity dispersion and
      the (red online) squares are the $3$D-velocity dispersion.}
    \label{Fig:NFWhalo}
  \end{center}
\end{figure}

\subsection{Cluster merger without DM}

\begin{table}
  \centering
  \caption{Properties of the merger objects in the no-DM simulations.
    The first column gives the SFE. The second, third and fourth
    columns show the parameters of a King profile fit to the resulting
    merger object, namely the core radius $r_{\rm c}$, the tidal radius
    $r_{\rm t}$ and the central surface brightness $\mu_{0}$ assuming a
    stellar mass to light ratio of $1.0$.  This choice is arbitrary,
    because the real value is unknown, but in agreement with visual M/L
    ratios of many old globulars \citep[e.g.][]{har96}.}
  \label{tab:res1}
  \begin{tabular}{cccc} \hline
    SFE & $r_{\rm c}$ & $r_{\rm t}$ & $\mu_{0}$ \\
    $[\%]$ & [pc] & [pc] & [mag\,arcsec$^{-2}$] \\ \hline
     100 &  7 & 50.1 & 21.3 \\
      33 & \multicolumn{3}{c}{dissolved} \\
      10 & \multicolumn{3}{c}{dissolved} \\ \hline
   \end{tabular}
 \end{table}

The first simulations we perform are of two star clusters which merge
under their own gravity without a DM halo present.  In
Table~\ref{tab:res1} we show the properties of the merger object after
$10$~Gyr of evolution.  This long time-span is chosen to ensure that
the object has had time enough to virialize and evolve slowly.  We fit
King profiles ~\citep{kin62} to the final merger objects for this and all models
studied because these profiles are widely used in observational astronomy and, in
particular, were used in the discovery paper of EGC Scl-dE GC1
~\citep{cos09}.  In reality, we would not expect a merger remnant to
exhibit a profile with a sharp tidal truncation radius, as in a King
profile.  Therefore the nominal value of the King tidal radius should
be regarded more as a third parameter of the fit rather than a
physical property of the remnant.

In the SFE$=100$~\% case, the fit gives a core radius $r_{\rm c}=7$~pc.
This is obviously a smaller value than the initial Plummer radius but
this does not imply that the final merger object is denser than its
two constituents.  Fitting a Plummer profile to the merger object
gives a fitted Plummer radius of about $11$~pc.  So the apparent
reduction in $r_{\rm c}$ only arises due to the different physical
meaning of the scale radius in the two profiles (King and Plummer).

If we introduce super-virial, expanding systems by increasing the
initial virial mass of the star cluster (which then gets artificially
reduced over one crossing-time), thereby mimicking a lower SFE
effciency of only $33$~\% (or $10$~\% respectively) and the expansion
due to the gas-expulsion, we do not find a merger object at all.  The
two star clusters simply disperse and build a very extended
distribution of stars, which in the presence of an external host
galaxy would have no chance to survive.

\begin{table}
  \centering
  \caption{Results of the simulations of cluster mergers in low mass
    haloes.  In all cases, the mass of the halo (enclosed within
    $500$~pc) is $5.6 \times 10^{4}$~$[$M$_{\odot}]$.  The first three
    columns indicate the profile of the DM halo (P for a Plummer
    profile, N for a NFW profile), the choice of initial cluster
    relative velocities (with notation according to Table~\ref{tab:vel}),
    and the adopted SFE.  Columns four to six show the parameters of
    the fitted King profile to the final merger object (i.e.\ core 
    radius $r_{\rm c}$, half-light radius $r_{\rm h}$ (for the fitted 
    King profile), tidal radius $r_{\rm t}$ and central surface brightness 
    $\mu_0$), while the
    seventh column gives the ratio of DM to luminous matter within the
    central $50$~pc of the object.  This can be transformed into a
    regular M/L ratio by adding the masses of both components and
    adopting a stellar M/L ratio for the luminous component.  For
    example, having a mass-to-mass ratio of unity and adopting a M/L
    for the luminous component of unity as well, we get a total M/L
    of 2.}
  \label{tab:res2}
  \begin{tabular}{cccccccc} \hline
    profile & velocity & SFE & $r_{\rm c}$ & $r_{\rm h}$ & $r_{\rm t}$ &
    $\mu_{0}$ & $\frac{M_{\rm DM}}{M_{\rm star}}$ \\
    P/N & case & [\%] & [pc] & [pc] & [pc] & [mag/ &
     \\ 
     & & & & & & arcsec$^{2}$] & \\ \hline
     P & 1 & 100 & 7 & 16.9 & 45 & 21.4 & 0.01 \\
     N & 1 & 100 & 3.5 & 11.1 & 31 & 29.4 & 0.01 \\
     P & 1 &  33 & 14 & 39.4 & 108 & 23.3 & 0.01 \\
     N & 1 &  33 & 24 & 50.9 & 132 & 24.5 & 0.039 \\
     P & 1 &  10 & \multicolumn{5}{c}{dissolved} \\
     N & 1 &  10 & \multicolumn{5}{c}{dissolved} \\
     P & 2 & 100 & 7 & 18.5 & 50 & 21.5 & 0.6 \\
     N & 2 & 100 & \multicolumn{5}{c}{did not merge} \\
     P & 2 &  33 & 15 & 35.4 & 94 & 23.3 & 0.13 \\
     N & 2 &  33 & 21.5 & 69.7 & 196 & 24.6 & 0.036 \\
     P & 2 &  10 & \multicolumn{5}{c}{dissolved} \\
     N & 2 &  10 & \multicolumn{5}{c}{dissolved} \\
     P & 3 & 100 & 7 & 19.7 & 54 & 21.6 & 0.06 \\
     N & 3 & 100 & \multicolumn{5}{c}{did not merge} \\
     P & 3 &  33 & 22 & 58.9 & 160 & 24.9 & 0.16 \\
     N & 3 &  33 & 26 & 53.9 & 139 & 24.6 & 0.02 \\
     P & 3 &  10 & \multicolumn{5}{c}{dissolved} \\
     N & 3 &  10 & \multicolumn{5}{c}{dissolved} \\
     P & 4 &  100 & 7 & 23.4 & 66 & 21 & 0.01 \\
     N & 4 &  100 & 3.5 & 12.9 & 37 & 19.4 & 0.01 \\
     P & 4 &  33 & 22 & 64.1 & 177 & 24.9 & 0.16 \\
     N & 4 &  33 & 21 & 41.5 & 106 & 24.6 & 0.13 \\
     P & 4 &  10 & \multicolumn{5}{c}{dissolved} \\
     N & 4 &  10 & \multicolumn{5}{c}{dissolved} \\ \hline
   \end{tabular}
 \end{table}
\begin{figure*}
  \centering
  \epsfxsize=7.0cm
  \epsfysize=7.0cm
  \epsffile{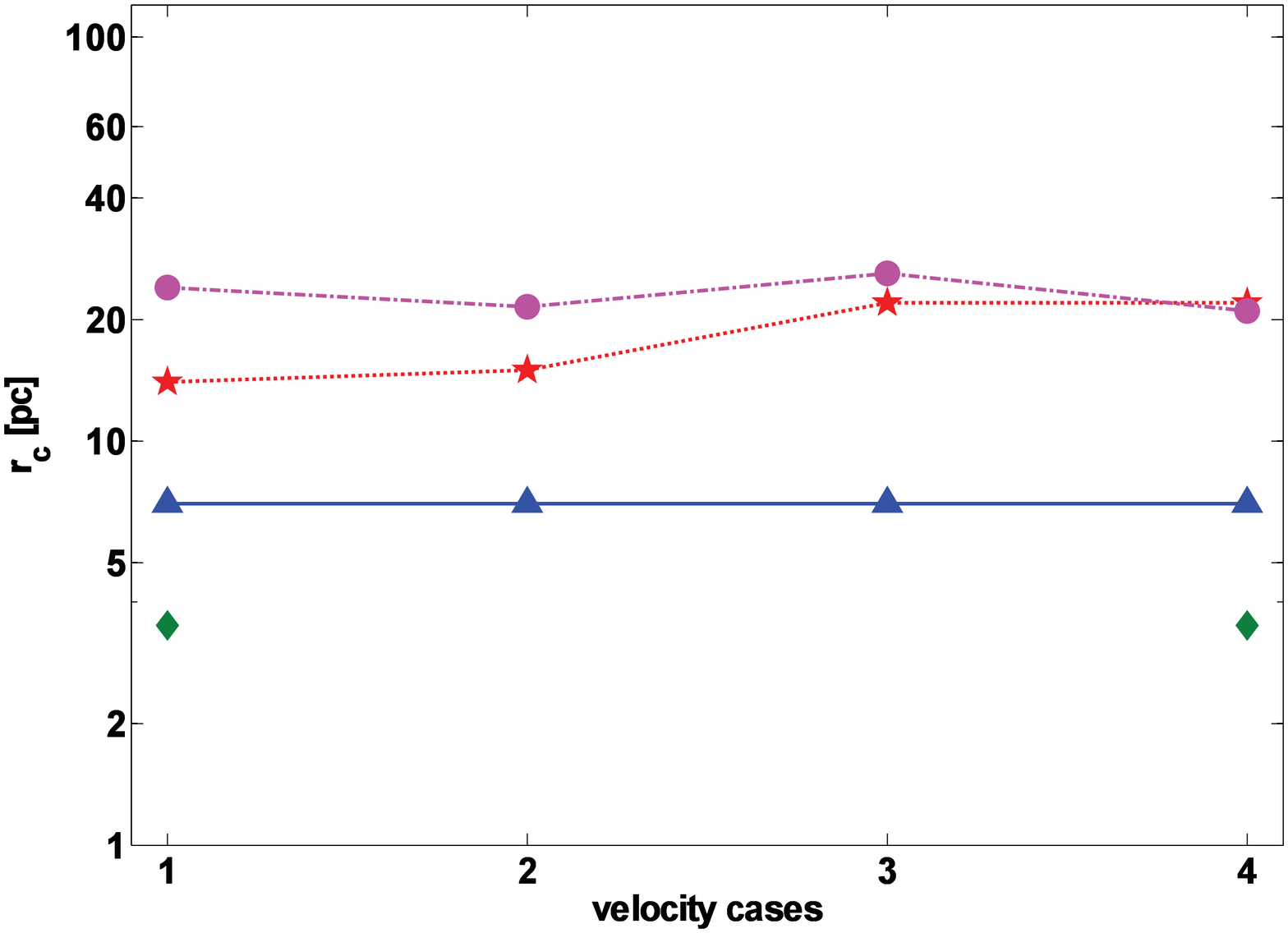}
  \epsfxsize=7.0cm
  \epsfysize=7.0cm
  \epsffile{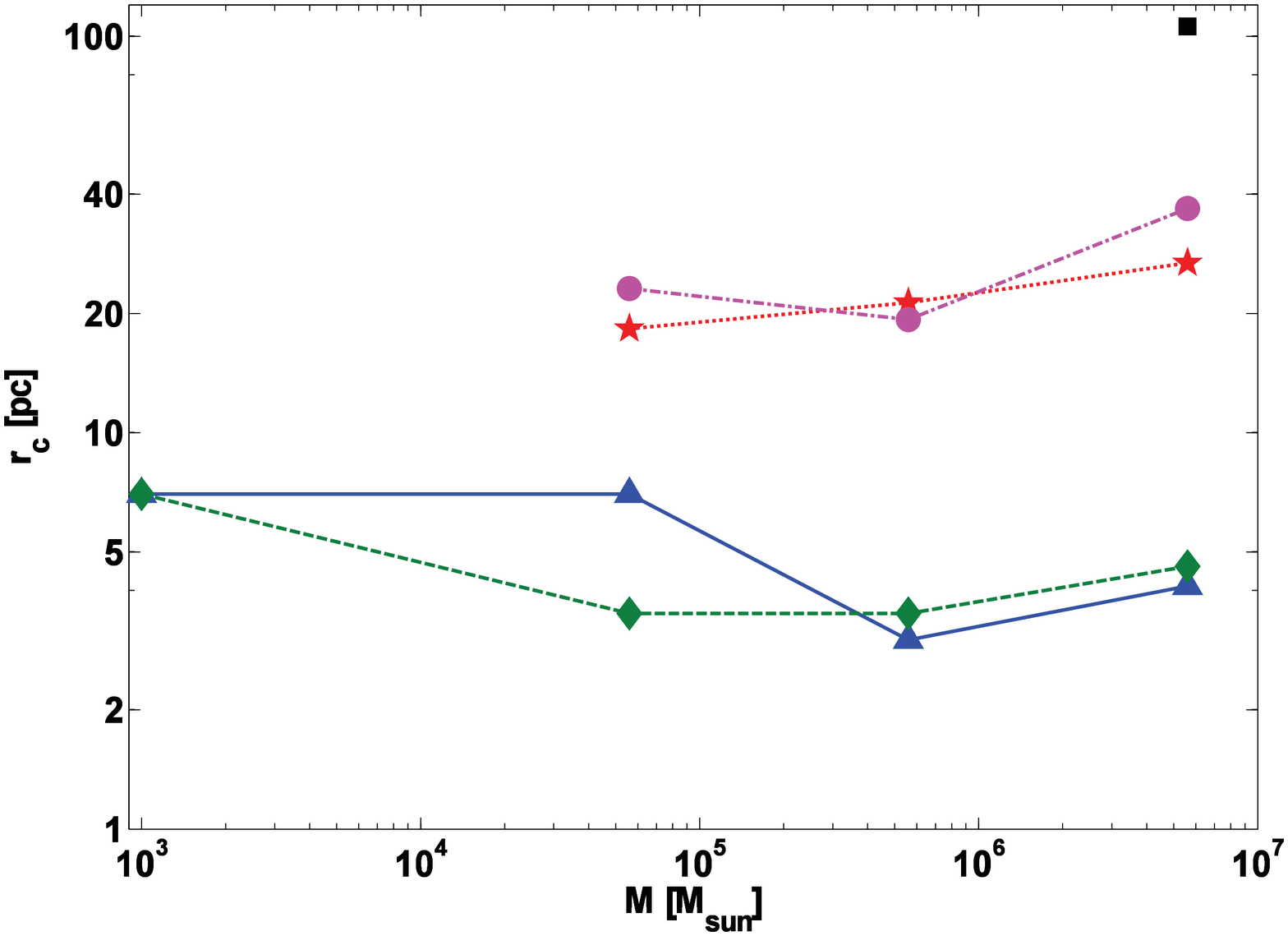}
  \epsfxsize=7.0cm
  \epsfysize=7.0cm
  \epsffile{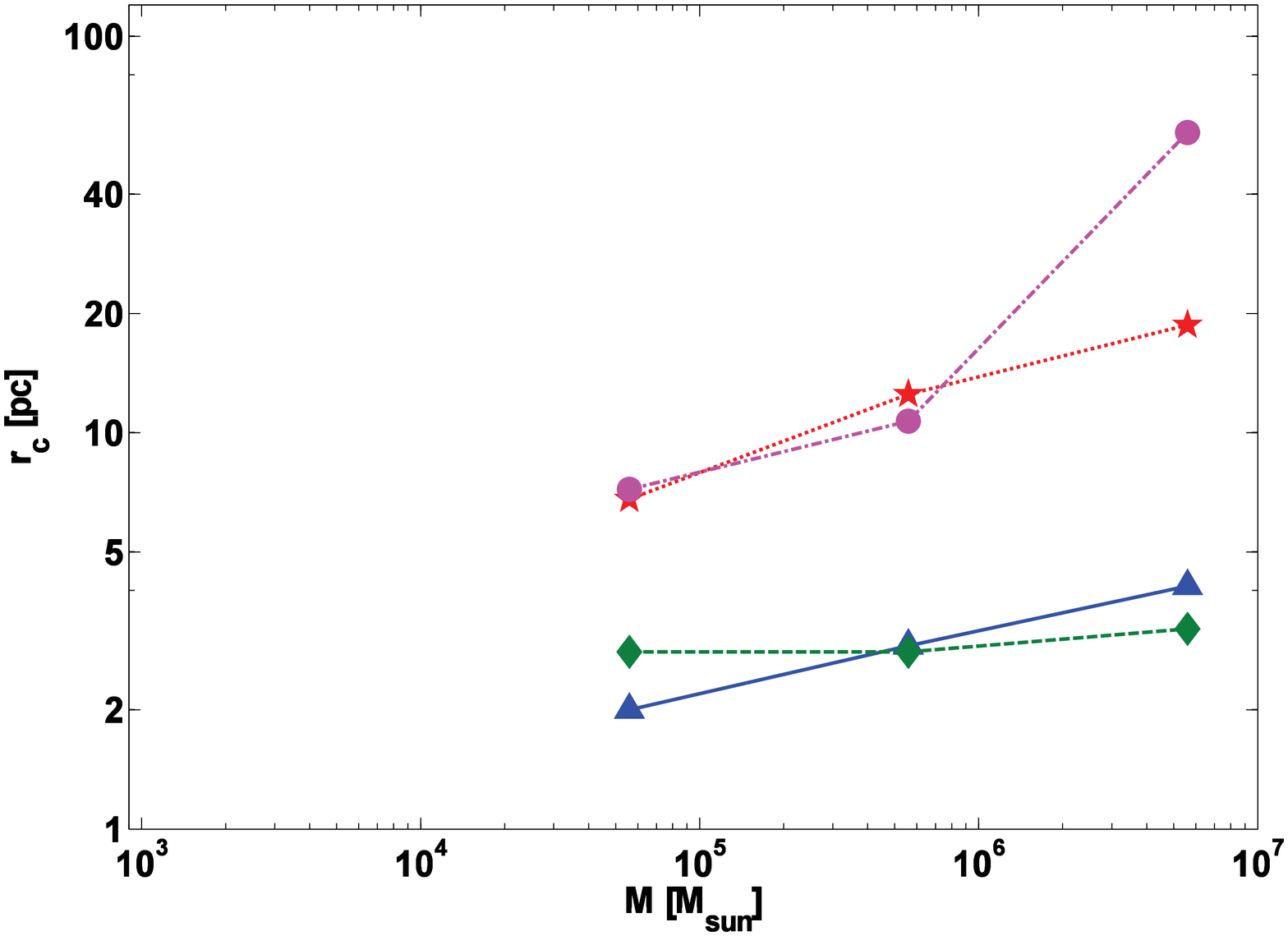}
  \epsfxsize=7.0cm
  \epsfysize=7.0cm
  \epsffile{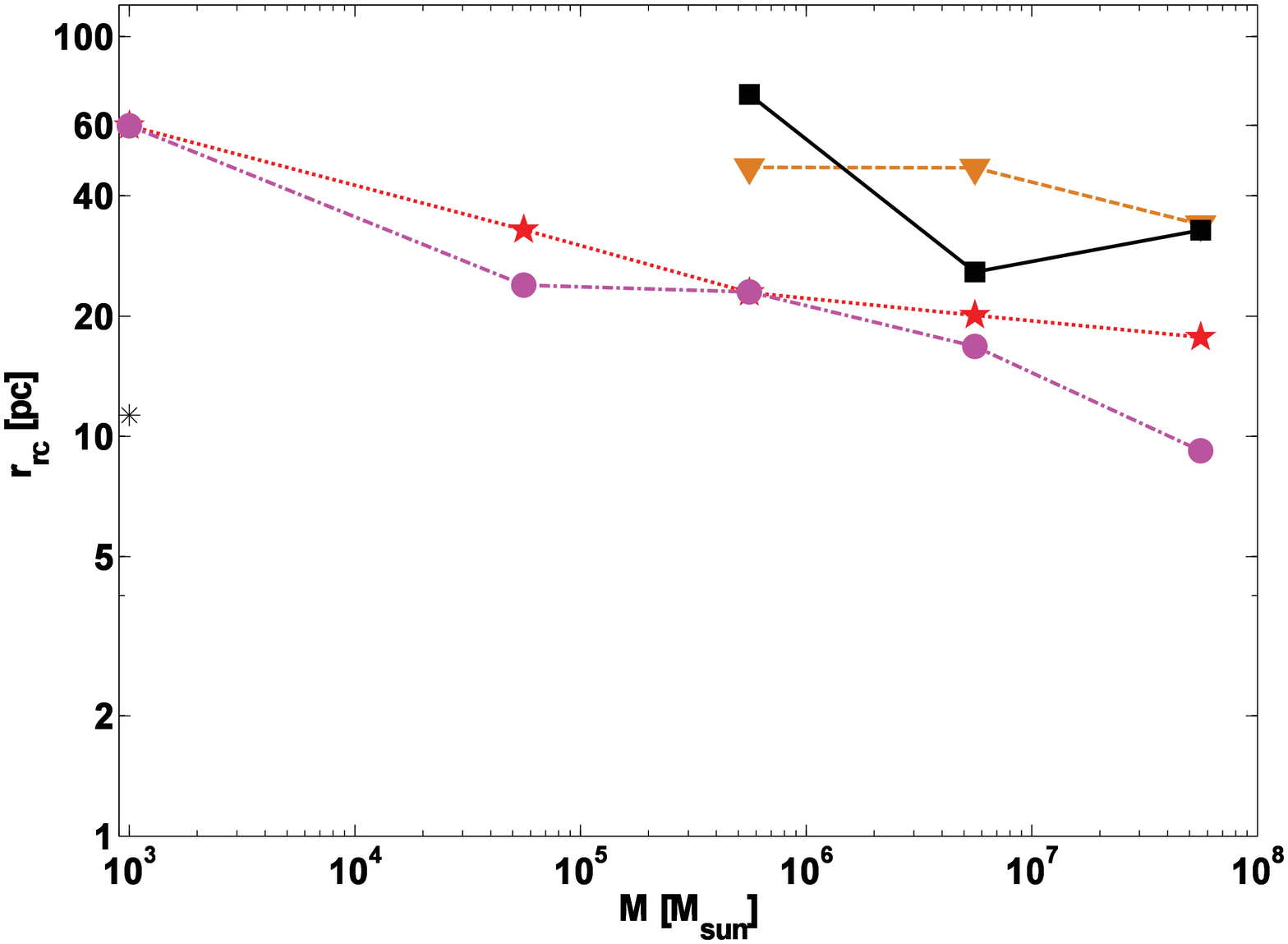}
  \caption{\textbf{Top-left:} Core radius of our merger objects as
    function of the adopted relative velocity case (see Table~\ref{tab:vel}).
    The halo mass was $5.6 \times 10^{4}$~M$_{\odot}$ in all cases.
    Triangles (blue online) show Plummer profiles and diamonds (green
    online) show NFW profiles for the halo with a SFE of $100$~\% for
    the two star clusters.  Stars (red online) and points (magenta
    online) show Plummer and NFW shapes for the halo, respectively, if
    the SFE is $33$~\%.  \textbf{Top-right:} The core radii of our
    merger objects as function of the mass of the halo.  Triangles
    (blue online) show Plummer haloes and star clusters with $100$~\%
    SFE.  Diamond (green online) the same SFE but with NFW haloes.
    Stars (red online) show Plummer haloes using $33$~\% for the
    SFE and points (magenta online) the corresponding NFW cases.  The
    results for the simulations without DM are for comparison shown at
    a halo mass of 1000 because of the logarithmic scale.  The only
    resulting merger object at a SFE of $10$ inside a NFW halo is
    shown as the square (black online). \textbf{Bottom-left:} Core
    radius versus halo mass for our models with compact star clusters.
    Triangles (blue) and Diamonds (green) show star clusters with $100$~\%
    SFE in haloes with Plummer and NFW profiles, respectively.
    Stars (red online) show Plummer type haloes and a star
    cluster with a SFE of $33$~\%.  Circles (magenta) show the same
    for haloes with NFW profiles. \textbf{Bottom-right:} Core radius
    versus halo mass for our single star cluster models.  Stars
    (red online) show Plummer type haloes and a star cluster with a
    SFE of $33$~\%.  Points (magenta) show the same for haloes with
    NFW profiles.  Square (black) and Triangles down (orange) show star clusters
    with $10$~\% SFE in haloes with Plummer and NFW profile,
    respectively. Black asterisk show the resulted star cluster from initial compact star cluster evolved without dark matter halo. See text for a detailed discussion.}
  \label{fig:rcore}
\end{figure*}

\subsection{Cluster merger with DM}

We now place both star clusters in a common, very low-mass DM halo
($M_{\rm DM}= 5.6\times 10^{4}~$M$_{\odot}$) and test for the
influence of the encounter geometry.  The results of these simulations
are given in Table~\ref{tab:res2}.  For the SFE$=100$~\% cases, the
core radii of the resulting EGCs are independent of the initial
relative velocities of the clusters, with Plummer profiles yielding
EGCs with $r_{\rm c} = 7$~pc while NFW haloes lead to more compact
EGCs with $r_{\rm c} = 3.5$~pc. We find two simulations, using NFW
haloes, in which the SCs do not merge and remain as two independent
structures.  These are the cases in which we give the star clusters
transverse velocities of the order of the halo circular velocity.  In
those cases the star clusters meet off-centre but are compact enough
not to merge.  This only happens in the NFW models because here the
cusp ensures that the orbits of both clusters never lead to close
encounters between the clusters.  Because of the low mass of the halo
and the low densities in the outer regions of the two clusters (we use
steep Plummer profiles as inital models), dynamical friction is not
strong enough to bring both clusters close enough to merge.  However,
EGCs always form if the SFE is $33$~\%.  Two of these simulations,
which have a Plummer DM halo, result in EGCs with $r_{\rm c}=22$~pc.
The EGC formed inside a NFW DM halo which most closely resembles the
properties of Scl-dE GC1 is formed when the relative velocities have a
small transversal component (case 2 in Table~\ref{tab:vel}), i.e.\ the
two star clusters do not merge in a head-on collision.  In the top
panel of Figure~\ref{Fig:NFWhalo} we show the surface density profile
of this cluster.

It is interesting to note that in all cases with a low SFE of $10$~\%,
we do not form a merger object at all but the two clusters instead
dissolve more quickly than they are able to merge. The result is a
system with a diffuse light distribution, more similar to a dSph
galaxy than a star cluster. In~\cite{ass10} and Assmann et al. (in
prep.), we explore this scenario for the formation of dSphs in detail.

In Figure~\ref{fig:rcore}, we show in the top-left panel the resulting
core radii as function of initial cluster relative velocity (see
Table~\ref{tab:vel}).  It shows clearly that the size of the merger
object depends mainly on the adopted SFE and not on the geometry of
the first encounter, as long as the two clusters are able to merge.
We find no significant general trend in our simulations.  As a minimal,
second order trend we see, that if the star clusters meet slightly
off-centre (velocity cases 2+3) the resulting core radii are mostly
slightly larger and if we increase the speed of the head-on encounter
(case 4) we get in most of the cases a smaller core radius.

 \begin{table}
   \centering
   \caption{Results of our simulations for different halo masses
     (enclosed within a radius of $500$~pc).  The first column gives
     the shape of the halo (- for none, P for Plummer and N for NFW),
     the second column the mass of the halo and the third the SFE.
     The fourth, fifth and sixth columns give the parameters of a
     King-profile fitted to the data, namely the core radius, tidal
     radius and central surface brightness.  The last column gives the
     mass ratio between the DM and the luminous matter within the
     central $50$~pc.  The results shown are mean values from up to
     four different realisations. }
   \label{tab:res3}
   \begin{tabular}{cccccccc} \hline
     profile & $M_{\rm h}$ & SFE & $r_{\rm c}$ & $r_{\rm h}$ & $r_{\rm t}$ & 
     $\mu_{0}$ & $\frac{M_{\rm DM}}{M_{\rm star}}$ \\
     P/N & [M$_{\odot}$ & [\%] & [pc] & [pc] & [pc] & [mag/ & \\ 
     & & & & & & \,arcsec$^{2}$] & \\ 
     \hline
     - & 0.0               & 100 &  7.0 & 18.5 & 50.1 & 21.3 & 0.0 \\
     - & 0.0               &  33 & \multicolumn{5}{c}{dissolved} \\
     \hline
     P & 5.6$\times 10^{4}$ & 100 &  7.0 & 18.4 & 49.7 & 21.5 & 0.025 \\
     N & 5.6$\times 10^{4}$ & 100 &  3.5 & 11.1 & 31.0 & 29.4 & 0.01 \\
     P & 5.6$\times 10^{4}$ &  33 & 18.3 & 49.6 & 135.0 & 24.1 & 0.01 \\
     N & 5.6$\times 10^{4}$ &  33 & 23.1 & 54.0 & 143.0 & 24.6 & 0.06 \\
     P & 5.6$\times 10^{4}$ &  10 & \multicolumn{5}{c}{dissolved} \\
     N & 5.6$\times 10^{4}$ &  10 & \multicolumn{5}{c}{dissolved} \\
     \hline
     P & 5.6$\times 10^{5}$ & 100 &  3.0 & 11.0 & 31.5 & 19.6 & 0.1 \\
     N & 5.6$\times 10^{5}$ & 100 &  3.5 & 12.3 & 35.2 & 19.6 & 0.13 \\
     P & 5.6$\times 10^{5}$ &  33 & 21.3 & 85.6 & 249.0 & 24.4 & 0.31 \\
     N & 5.6$\times 10^{5}$ &  33 & 19.3 & 103.7 & 313.0 & 24.7 & 0.06 \\
     P & 5.6$\times 10^{5}$ &  10 & \multicolumn{5}{c}{dissolved} \\
     N & 5.6$\times 10^{5}$ &  10 & \multicolumn{5}{c}{dissolved} \\
     \hline
     P & 5.6$\times 10^{6}$ & 100 &  4.1 & 9.5 & 25.2 & 21.1 & 1.06 \\
     N & 5.6$\times 10^{6}$ & 100 &  4.6 & 29.6 & 91.0 & 20.8 & 3.37 \\
     P & 5.6$\times 10^{6}$ &  33 & 26.8 & 45.4 & 112.0 & 23.6 & 1.69 \\
     N & 5.6$\times 10^{6}$ &  33 & 36.8 & 596 & 1950  & 25.8 & 24.4 \\
     P & 5.6$\times 10^{6}$ &  10 & \multicolumn{5}{c}{dissolved} \\
     N & 5.6$\times 10^{6}$ &  10 & 106 & 198 & 500 & 26.0 & 40.8 \\ \hline
     P/N & 5.6$\times 10^{7}$ &  \multicolumn{6}{c}{all simulations
       dissolve} \\ \hline
     P/N & 5.6$\times 10^{8}$ &  \multicolumn{6}{c}{all simulations
       dissolve} \\ \hline
  \end{tabular}
\end{table}

As a next step, we increase the mass of the halo by factors of $10$.
The results of our simulations are given in Table~\ref{tab:res3}.
Again if the SFE is $100$~\% we only obtain merger objects with core
radii of only $r_{\rm c}=3$--$7$~pc independent of the mass of the
halo.  As soon as the SFE is $33$~\%, it is possible to obtain an EGC
which resembles Scl-dE1 GC1 as long as the halo does not become too
massive, i.e.\ the DM content within the final merger object is well
below its luminous mass.  However, it is interesting to note that in
the two cases with $M_{\rm h} = 5.6 \times 10^{5}$~M$_{\odot}$ the
EGCs have a concentration ($c \equiv \log10(r_{t}/r_{\rm c})$) that is
double the estimated concentration of $0.65$ for EGC Scl-dE GC1
reported by ~\citep{cos09}.

If we consider a mass for the DM halo of $M_{h}=5.6 \times
10^{6}$~M$_{\odot}$ and a SFE of $33\%$, the simulated EGCs have core
radii which are too large ($r_{\rm c}=26-39$~pc).  In other words when
the DM content within the merger object is of the same order as that
of the luminous matter (mass ratio of $1.7$) or significantly higher
(mass ratio of $24.4$) we get merger objects which are too extended.
However, in those cases it may be possible to find a matching model by
tuning the SFE to higher values. In almost all simulations with a SFE
of $10$~\%, the SCs dissolve before they merge.

Simulations using a higher mass for the DM halo ($5.6 \times
10^{7}$~$[$M$_{\odot}]$ and $5.6 \times 10^{8}$~$[$M$_{\odot}]$
enclosed within $500$~pc), lead to completely dissolved star clusters
independent of the SFE used.  In none of these simulations did we
actually see the merging of the two clusters.  This can be understood
using the theory developed in \cite{fel09}.  If the strength of the
background potential becomes too high, i.e.\ the star clusters meet
each other with relative velocities much higher than their internal dispersions
it results in destructive high-speed encounters rather than in low
velocity merging.

In the top-right panel of Figure\ref{fig:rcore} we plot the core radius
of our merger objects as function of the halo mass.  We clearly see
that the mass of the halo has only a small influence on the final core
radius of the resulting merger object, except for the fact that we
have no merger object in very high mass haloes.  The main parameter
which governs the results is clearly the SFE of the two star clusters.
This SFE has to be of order $33$~\% to obtain a merger object which
resembles the EGC.

\cite{cos09} discussed whether the EGC Scl-dE GC1 could be a dark
matter dominated object.  Our most important result is that we only
get objects which resemble this EGC if the mass of the DM halo within
the final stellar distribution is negligible.  Higher halo masses lead
either to objects which are too extended (i.e. we have to use higher
SFEs to counteract external destruction) or to no merged objects at
all. {\color{red}To understand this behavior we need to consider all
  the competing processes which drive the evolution of the merging
  clusters within the DM halo. First, it is important to observe that
  the expansion of the star clusters is governed by the SFE, such that
  lower SFEs give more extended objects \citep{baum07}. Secondly, it
  must be emphasized that the influences of the DM halo can be divided
  into two distinct processes, that depend on where the SCs are
  initially located. One process is when the star clusters have not
  yet merged in the centre of the DM halo. In this case, the tidal
  forces generated by DM halo act as an additional destructive force
  on the clusters, reducing their masses and sizes. The more massive
  the DM halo, the more dominant is this disruptive
  process. Additionally, a higher halo mass increases the final
  relative velocities with which the SCs meet and might, therefore,
  inhibit the merging. This first halo process tends to lead to
  remnants which are more extended and diffuse than observed EGCs. The
  second halo process takes place when the merged object forms and
  remains at rest in the centre of the DM halo. In this case, the
  potential well of the DM halo contributes to the gravitational field
  in which the cluster stars move. As a result, objects which might
  otherwise dissolve are bound by the DM halo, and the additional
  gravitational field makes the remnants more compact than they would
  be without the presence of DM.}

Our simulations suggest that mass-to-light
ratios of only a few would be expected if the object is formed by the
merger of two star clusters, and thus, they are consistent with there being
no dynamically significant amount of DM within the stellar distribution
of the EGC.  However, our results also show that we do require some DM
to ensure that the clusters merge to produce an EGC-like remnant, so
that DM is an essential part of the merger scenario per se despite not
being a significant component of the merger remnant.

\begin{table}
  \centering
  \caption{The central line-of-sight velocity dispersions of our
    best-fitting models, i.e.\ whose $r_{\rm c}$ is close to that of
    Scl-dE GC$1$.  All cases shown here have a SFE of $33$~\%.  For a
    halo mass of $5.6 \times 10^{4}$M$_{\odot}$ column three gives the
    velocity case, while for halo masses of
    $5.6 \times 10^{5}$~M$_{\odot}$ we give the realisation number.}
  \label{tab:vel2}
  \begin{tabular}{ccccc} \hline
    $M_{\rm halo}$ & profile & velocity & $r_{\rm c}$ &
    $\sigma_{\rm LOS,mean}$  \\
    $[$M$_{\odot}]$ & P/N & case & pc & km\,s$^{-1}$
    \\ \hline
    $5.6 \times 10^{4}$& N & $2$ & $21.5$ & $0.78$ \\
    $5.6 \times 10^{4}$& P & $3$ & $22$ & $0.94$ \\
    $5.6 \times 10^{4}$& P & $4$ & $22$ & $1.17$ \\
    $5.6 \times 10^{4}$& N & $4$ & $21$ & $1.68$ \\ \hline
    $5.6 \times 10^{5}$& P & $1$ & $21.4$ & $1.63$ \\
    $5.6 \times 10^{5}$& P & $2$ & $23$ & $1.66$ \\
    $5.6 \times 10^{5}$& N & $2$ & $19.5$ & $1.66$ \\
    $5.6 \times 10^{5}$& P & $3$ & $19.6$ & $1.65$ \\
    $5.6 \times 10^{5}$& N & $3$ & $19.5$ & $1.62$ \\
    $5.6 \times 10^{5}$& P & $4$ & $21$ & $1.64$ \\
    $5.6 \times 10^{5}$& N & $4$ & $20$ & $1.61$ \\ \hline
  \end{tabular}
\end{table}

In addition to comparing the core radius and concentration of our
merger remnants with those of EGC Scl-dE GC1, we can also use our
simulations to look at the velocity space of the EGC.
Table~\ref{tab:vel2} shows the mean, central line-of-sight velocity
dispersion ($\sigma_{\rm LOS,mean}$) for the simulations which
resemble the EGC.  We average the central dispersion measured along
all three coordinates axes (the two coordinates which span the plane
of the interaction and the one perpendicular to that) because the
orientation of the sight-line towards GC1 relative to any proposed
merger is, of course, unknown. In our models, $\sigma_{\rm LOS,mean}$
varies from $0.78$~km$s^{-1}$ to $1.68$~km\,$s^{-1}$.  In
Figure~\ref{Fig:NFWhalo} we show the dispersion profile for one of our
merger objects.  It is notable that it is possible to increase the
central dispersion of our merger object up to almost
$1.7$~km\,s$^{-1}$ without having significant amounts of DM within the
object.

The importance of studying the velocity space of the simulated
clusters is that the velocity dispersions can be used to distinguish
between our various scenarios for the formation of EGC Scl-dE GC1.
Moreover, the velocities obtained in our simulations can be compared
with future observations to infer the presence or absence of dark
matter in the system.

\subsection{Compact initial clusters}
\label{sec:4pc}

For comparison, we also perform simulations with a more standard model
for the initial star cluster, i.e.\ with a Plummer radius of just
$4$~pc.  With this suite of simulations we want to investigate whether
it is possible to obtain EGCs like GC1 without starting with star
clusters that are already extended.

\begin{table}
  \centering
  \caption{Results of our simulations for star clusters which
    are initially more concentrated, i.e.\ have a Plummer radius of
    $4$~pc.  The columns are the same as in Table~\ref{tab:res3}.}
  \label{tab:res5}
  \begin{tabular}{cccccccc} \hline
    profile & $M_{\rm halo}$ & SFE & $r_{\rm c}$ & $r_{\rm h}$ & $r_{\rm t}$ &
    $\mu_{0}$ & $\frac{M_{\rm DM}}{M_{\rm star}}$ \\
    P/N & [M$_{\odot}$] & [\%] & [pc] & [pc] & [pc] & [mag/ &
    \\ 
    & & & & & & \,arcsec$^{2}$] & \\ \hline
    --- & --- & 100 & \multicolumn{5}{c}{did not merge} \\
    --- & --- &  33 & \multicolumn{5}{c}{did not merge} \\
    \hline
    P& $5.6 \times 10^{4}$ & 100 & 2 & 25.1 & 81 & 19.2 & 0.03  \\
    N& $5.6 \times 10^{4}$ & 100 & 2.8 & 12.5 & 37 & 19.1 & 0.01 \\
    P& $5.6 \times 10^{4}$ & 33 & 6.8 & 12.1 & 30.3 & 22.05 & 0.06 \\
    N& $5.6 \times 10^{4}$ & 33 & 7.2 & 24.1 & 68.2 & 22.0 & 0.02 \\
    P& $5.6 \times 10^{4}$ & 10 & \multicolumn{5}{c}{dissolved} \\
    N& $5.6 \times 10^{4}$ & 10 & \multicolumn{5}{c}{dissolved} \\
    \hline
    P& $5.6 \times 10^{5}$ & 100 & 2.9 & 9.9 & 28 & 19.3 & 0.13  \\
    N& $5.6 \times 10^{5}$ & 100 & 2.8 & 14.0 & 42 & 19.4 & 0.2 \\
    P& $5.6 \times 10^{5}$ & 33 & 12.5 & --- & 1.2e+5 & 24.2 & 0.35 \\
    N& $5.6 \times 10^{5}$ & 33 & 10.7 & --- & 1.2e+3 & 23.9 & 0.7 \\
    P& $5.6 \times 10^{5}$ & 10 & \multicolumn{5}{c}{did not merge} \\
    N& $5.6 \times 10^{5}$ & 10 & \multicolumn{5}{c}{dissolved} \\
    \hline
    P& $5.6 \times 10^{6}$ & 100 & 4.1 & 10.4 & 28 & 21.88 & 1.1  \\
    N& $5.6 \times 10^{6}$ & 100 & 3.2 & 33.2 & 106 & 20.33 & 3.34 \\
    P& $5.6 \times 10^{6}$ & 33 & 18.7 & --- & 3.1e+4  & 24.5 & 0.35 \\
    N& $5.6 \times 10^{6}$ & 33 & 57.2 & --- & 1.2e+5 & 26.1 & 24.1 \\
    P& $5.6 \times 10^{6}$ & 10 & \multicolumn{5}{c}{dissolved} \\
    N& $5.6 \times 10^{6}$ & 10 & \multicolumn{5}{c}{dissolved} \\
    \hline
  \end{tabular}
\end{table}

The results in Table~\ref{tab:res5} and the bottom-left panel of
Figure~\ref{fig:rcore}, show clearly that we do not end up with a
merger object that resembles the properties of GC1, if the initial star
clusters are too concentrated.  A SFE which guarantees the
survival of the two clusters always leads to merged objects which are
too concentrated to resemble the EGC in Sculptor.  All our merged
objects show core radii of less than or about $10$~pc. This behavior is
explained by the force performed by the DM halo as soon as
the merger SCs are settled in the centre. This force keeps the merged object
together and more compact. The exceptions are the simulations with
a $5.6 \times 10^{6}$~M$_{\odot}$ halo. Here again we see a completely dissolved
object but this time it is possible to determine a centre of density.
However, even though the fitted 'core radii' seem to match the properties of the EGC,
one sees from the values of the 'tidal radii' that these are not really EGCs but
rather fluffy, low-density distributions of stars.  Taking these results
into account we can rule out that the EGC has formed out of two
concentrated star clusters.

\subsection{Abandoning the merger scenario}
\label{sec:1clus}

To complete our survey of possible formation scenarios we now abandon
the hypothesis that Scl-dE1~GC1 formed via a merger of multiple
clusters and explore whether it is possible to expand a single young
star cluster sufficiently due to gas-expulsion to resemble the
observed EGC.

\begin{table}
   \centering
   \caption{Results of our simulations with a single star cluster
     expanding because of mass-loss due to gas-expulsion. The columns
     are the same as in Table~\ref{tab:res3}.  For the non-DM
     simulations we give two simulations which have a SFE of $33$~\%.
     The first simulation started with a more compact Plummer sphere
     and has a Plummer radius of $4$~pc, while the second one shows
     the standard $11$~pc model.}
  \label{tab:res6}
  \begin{tabular}{ccccccccc} \hline
    profile & $M_{\rm h}$ & SFE & $r_{\rm c}$ & $r_{\rm h}$ & $r_{\rm t}$ &
    $\mu_{0}$ & $\frac{M_{\rm DM}}{M_{\rm star}}$ \\
    P/N & [M$_{\odot}$] & [\%] & [pc] & [pc] & [pc] & [mag/ &
      \\ 
      & & & & & & \,arcsec$^{2}$] & \\ \hline
      - & 0.0 & 33 & 11.3 & 21.1 & 53.4 & 24 & ---   \\
      - & 0.0 & 33 & 59.9 & 62.7 & 139 & 27.6 & ---   \\
      - & 0.0 & 10 & \multicolumn{5}{c}{dissolved} \\
      \hline
      P & $5.6 \times 10^{4}$ & 33 & 32.9 & 34.5 & 76.5 & 23.6 & 0.23   \\
      N & $5.6 \times 10^{4}$ & 33 & 23.9 & 42.3 & 105.6 & 24.2 & 0.03 \\
      P & $5.6 \times 10^{4}$ & 10 & \multicolumn{5}{c}{dissolved} \\
      N & $5.6 \times 10^{4}$ & 10 & \multicolumn{5}{c}{dissolved} \\
      \hline
      P & $5.6 \times 10^{5}$ & 33 & 22.9 & 39.2 & 97.2 & 23.7 & 0.21 \\
      N & $5.6 \times 10^{5}$ & 33 & 23 & 34.3 & 82.3 & 23.3 & 0.35 \\
      P & $5.6 \times 10^{5}$ & 10 & 47.1 & 84.0 & 210 & 26.3  & 0.55 \\
      N & $5.6 \times 10^{5}$ & 10 & 71.7 & 67.7 & 147 & 25.7 & 0.5 \\
      \hline
      P & $5.6 \times 10^{6}$ & 33 & 20.1 & 37.9 & 96 & 23.1 & 0.8   \\
      N & $5.6 \times 10^{6}$ & 33 & 16.8 & 29.1 & 72.2 & 22.5 & 4.1 \\
      P & $5.6 \times 10^{6}$ & 10 & 47 & 51.4 & 115 & 24.1 & 1.34 \\
      N & $5.6 \times 10^{6}$ & 10 & 25.8 & 86.0 &  243 & 24.1 & 4.2 \\
      \hline
      P & $5.6 \times 10^{7}$ & 33 & 17.7 & 29.6 & 73 & 22.5 & 5.89 \\
      N & $5.6 \times 10^{7}$ & 33 & 9.2 & --- & 3.5e+4 & 23.8 & 74.8 \\
      P & $5.6 \times 10^{7}$ & 10 & 33.9 & 36.3 & 81 & 23.01 & 7.78 \\
      N & $5.6 \times 10^{7}$ & 10 & 32.8 & --- & 4.7e+5 & 25.34 & 91.2 \\
      \hline
    \end{tabular}
\end{table}

Our simulations (results are shown in Table~\ref{tab:res6}) show
clearly that it is possible to get an extended star cluster just by
expansion due to gas-expulsion alone.  In our simulations we do not
include any tidal field of the parent galaxy and therefore our star
cluster can expand freely without fearing destruction due to external
tidal forces.  In the case without DM, we use only our standard SFE of
$33$~\%, which has been shown in many studies~\citep[e.g.][]{baum07}
to be the lowest limit for a cluster to survive gas-expulsion.
Therefore it is no wonder that the cluster expands significantly.
However, by using higher SFE and/or initially more compact clusters it
is likely that we could find a suitable model to reproduce the
observed data.  {\color{red}We performed one simulation which started
  from a compact star cluster with a Plummer radius of only $4$~pc and
  was transformed into an EGC with a scale length of $11.3$~pc, while
  a cluster with an initial Plummer radius of $11$~pc was transformed
  into an EGC with a scalelength of $\sim60$~pc.  This clearly shows
  that we would be able to get a model which reproduces the observed
  data by using an intermediate initial scale-length.}
Applying Occam's Razor, this might be the simplest and most plausible
way to explain the origin of the EGC in Sculptor.  However, the
intention of this paper is mainly to test and investigate the
alternative, and more complicated, formation scenarios which have been
proposed in the literature.

If we place the single cluster in the centre of its own DM halo, we
first see that the more massive the halo gets the more likely it is
for the star cluster to survive even for SFEs as low as $10$~\%.  The
higher the halo mass the more gravitational potential there is to halt
the expansion and therefore the more compact our resulting object is.
This trend is visible in the bottom-right panel of
Figure~\ref{fig:rcore}.  With our initial cluster model and values for
the SFE we get the best matches for intermediate mass haloes of $5.6
\times 10^{4}$ to $5.6 \times 10^{6}$~M$_{\odot}$ for both kinds of
halo profile.  Again we see no significant trend with halo profile
except that we expect the NFW haloes to have more mass in the region
of interest than the equivalent Plummer model. However, from the
general behavior of our results we expect to be able to find a
suitable solution for any halo mass.  If we had extended our parameter
space to include even more massive haloes then we would likely have
obtained a match even for SFE of $10$~\%.  At lower halo masses we
could use either initially more compact clusters or assume higher
SFEs.

\begin{table}
  \centering
  \caption{The central line-of-sight velocity dispersion of our
    best-fitting, single-cluster models, i.e.\ whose $r_{\rm c}$ is
    close to the one of Scl-dE GC$1$.  All cases shown here have a SFE
    of $33$~\%.  {\color{red}In the first two lines we give the results obtained
    in our non-DM simulations. The values of $r_{\rm c}$ in these
    models bracket that of the observed EGC - as discussed in the
    text, we expect that a good model of the observed data could be
    obtained for an intermediate value of the initial cluster Plummer
    radius.}}
  \label{tab:vel3}
  \begin{tabular}{ccccc} \hline
    $M_{\rm halo}$ & profile & $r_{\rm c}$ & $\sigma_{\rm LOS,mean}$  \\
    $[$M$_{\odot}]$ & P/N & pc & km\,s$^{-1}$ \\
    \hline
    0 & --- & $11.3$ & $1.08$ \\
    0 & --- & $59.9$ & $0.18$ \\
    \hline
    $5.6 \times 10^{4}$& N & $23.9$ & $1.06$ \\
    $5.6 \times 10^{5}$& N & $23$ & $1.73$ \\
    $5.6 \times 10^{5}$& P & $22.9$ & $1.66$ \\
    $5.6 \times 10^{6}$& P & $20.1$ & $2.27$ \\
    \hline
  \end{tabular}
\end{table}

Finally in Table~\ref{tab:vel3} we show the LOS velocity dispersion of
our best-fitting, single-cluster models.  As expected, the central
velocity dispersion rises with DM halo mass, from below one
km\,s$^{-1}$ for the non-DM case up to more than two km\,s$^{-1}$ for
the DM dominated one,  and therefore will be the crucial parameter to
decide whether or not this very extended object has its own DM halo,
once spectroscopic data are available.

\section{Conclusions}
\label{sec:conc}

In this paper, we explored several of the proposed formation scenarios
for the recently-discovered, extended globular cluster GC1 associated
with the dwarf elliptical galaxy Scl-dE1 in the Sculptor group of
galaxies.  Even though extended globular clusters have been found in
the haloes of many galaxies, this is the first such object found in
orbit around a dwarf galaxy and exhibiting a very large core radius.

There are two obvious scenarios for the formation of an extended
stellar system starting from one, or several, individual constituents.
The first, and probably more natural, scenario is via the expansion
induced by either gas expulsion or stellar mass loss during the early
evolution of a normal star cluster. The other viable scenario is that
extended objects form via the merger of smaller systems.  A third,
more recent proposal for the formation of globular clusters is that
they might reside in their own DM halo, thereby helping to solve the
missing satellite problem.  We have extended those speculations to the
object of our study. In this paper, we have combined aspects of these
three scenarios in order to investigate thoroughly which initial
conditions can lead to an object which resembles the observations.

Taking all our merger simulations into account, we see the following
trends:

\begin{enumerate}
\item As expected, for an assumed SFE of 100\%, mergers of two
  clusters without the presence of DM produce a remnant which is of
  similar density to the original clusters, while lower SFEs lead to
  the dissolution of both initial clusters without merging. Finely
  tuned initial conditions would be required to produce a merger
  remnant via this route which matches Scl-dE1 GC1 in terms of size.
\item Mergers of two virialised clusters within a DM halo do not, in
  general, lead to a remnant which is more extended than the original
  clusters. This result is independent of the assumed halo mass.  We
  therefore need to invoke additional expansion either due to a SFE
  which is lower than $100$~\% or due to mass loss from early stellar
  evolution to produce extended remnants.
\item For high halo masses, we find that low SFEs can lead to the
  dispersal of the star clusters before they have time to merge. In
  those cases we see an extended luminous component which more closely
  resembles a dSph galaxy than a globular cluster.  This coincides
  with the formation theory for dSph galaxies we discuss
  in~\citet{ass10} and Assmann et al. (in prep.).
\item In all the simulations which led to an extended remnant whose
  appearance matched that of Scl-dE1 GC1, the mass of DM enclosed
  within the volume probed by the stars of the EGC was not sufficient
  to raise the mass to light ratio significantly above the stellar
  M/L.  Therefore, even though our simulations can not rule out that
  this extended globular cluster resides in its own DM halo we do not
  expect it to be a highly DM dominated object at the present day.
\item We have shown that our models can be tested by future
  observations. In our remnants, the higher the DM content, the higher
  is the central velocity dispersion of our merger object ranging from
  $0.7$ to $1.7$~km\,s$^{-1}$.  {\color{red}The next generation
    telescopes, for example the Extremely Large Telescope, may be able
    to measure such 
  differences either from the integrated spectra of the clusters or
  from observations of individual cluster stars, which can then be
  compared with the predictions of the models we have simulated.}
\item If we merge clusters which are initially as compact as typical
  globular clusters, we find that it is very difficult to form an
  object resembling the observed EGC.  We need to lower the SFE below
  the general survival limit while still requiring that the clusters
  merge in the centre of the halo before they dissolve
  completely. Such chance mergers are possible and have been found in
  other studies as well ~\citep{fel05}, but are unlikely to be a
  dominant formation avenue.
\end{enumerate}

Abandoning the merger scenario, we have shown that we can also form an
object like GC1 via the expansion due to gas-expulsion during the
formation of the cluster. In these models, we start with an object
forming in the very centre of a DM halo and show that gas expulsion
produces remnant clusters whose extent depends on the mass of the halo
- the higher the halo mass the more compact is the remnant.  With high
halo masses it is possible to get surviving and matching objects by
using SFEs which would lead to the complete dissolution of the cluster
in the absence of DM. Starting out with a more compact object leads to
a final remnant which is more compact.  Thus, for any given halo mass,
it is possible to match the data either by changing the SFE or by
changing the initial size of the object. Again, as in the merger
scenario, the observable difference between the models which resemble
Scl-dE1 GC1 is their velocity dispersion.  The higher halo mass,
and therefore the DM content within the EGC, the higher its velocity
dispersion will be.

{\color{red}
The velocity dispersions of globular clusters containing
Intermediate-Mass Black Holes (IMBH) at their centres have been
discussed by \citet{baum05} using numerical models.  The velocity
dispersion for a star cluster of similar stellar mass to our EGC is
raised by only $0.1$-$0.5$~km $s^{-1}$ in the inner $10$~per cent of
the half-mass radius.  The same would also be true for the stronger
effect of a binary IMBH.  In both cases, the increase in velocity
dispersion is seen only in the core the star cluster.  Other numerical
simulations show, that ebven if there is a high binary black hole
fraction within a globular cluster \citep{mac08}, the central line of
sight velocity dispersion is about $1.0$-$1.5$ km\,$s^{-1}$ about
double the value, but shows the regular fall-off at the outer radii as
well.  

To distinguish between the presence of a black hole and the presence
of dark matter distributed within the cluster would require the
measurement of the velocity dispersion as a function of radius within
the cluster.  If the EGCs contain dark matter, then their velocity
dispersions will remain higher than expected at large radii, not
solely at the centre.  For some EGcs, such observations may be
feasible with the next generation of extremely large telescopes.}

Finally, we note that it is possible to obtain an object like Scl-dE1
GC1 just by using expansion due to mass-loss (gas expulsion and/or
stellar winds) within a single young cluster.  Fine-tuning the initial
concentration of the cluster and, in our case, the SFE leads to an
object which matches the data.  In the light of Occam's razor this
might be the most simplest and straightforward solution to explain the
properties of Scl-dE1 GC1. The key test of this model would be the
survivability of the remnant in the tidal field of the host
galaxy. This would require further simulations of the evolution of the
cluster in the external tidal field following the period of expansion
and is beyond the scope the present paper.

We conclude that the unusually large size of the extended star cluster
Scl-dE1 GC1 does not require the presence of dynamically significant
DM in this system. We have shown that there are several formation
paths for this object involving mergers of star clusters in low-mass
dark matter haloes which result in a remnant that contains very little
DM interior to the stellar distribution. It is also possible to form
the object through the expansion of a single star cluster as a result
of mass loss. Thus, we have demonstrated that the formation scenarios
proposed for this system in the literature are plausible. To make
further progress requires the measurement of a precise velocity
dispersion for the object, as this would enable us to distinguish
between formation models which require DM and those which require only
baryonic matter.  To distinguish between scenarios which involve
merging of two or more initial objects and models which only expand
due to gas-expulsion alone, one might also be able to look for
irregular chemical abundances in the stars of the cluster which could
point to an origin in different clusters.  \\

{\bf Acknowledgments:} PA is supported through a CONICYT PhD
scholarship and wishes to thank W. Dehnen for his help with the NFW
profiles.  PA also announces help through the MECESUP travel grant
FSM0605.  MF acknowledges financial support through FONDECYT grant
no. 1095092.  MIW acknowledges the Royal Society for support through a
University Research Fellowship.

\bibliographystyle{mnras}

\label{lastpage}

\end{document}